# Deconvolution by simulation

## Colin Mallows[1]

*Avaya Labs*

**Abstract:** Given samples $(x_1, \ldots, x_m)$ and $(z_1, \ldots, z_n)$ which we believe are independent realizations of random variables $X$ and $Z$ respectively, where we further believe that $Z = X + Y$ with $Y$ independent of $X$, the problem is to estimate the distribution of $Y$. We present a new method for doing this, involving simulation. Experiments suggest that the method provides useful estimates.

## 1. Motivation

The need for an algorithm arose in work on estimating delays in the Internet. We can send a packet from an origin A to a remote site B, and have a packet returned from B to A; the time that that this takes is called the "round-trip delay" for the link A-B. These delays are very volatile and are occasionally large. We can also send packets from A to a more remote site C, by way of B, and can arrange for packets to be returned from C via B to A; this gives the round-trip delay for the A-B-C path. However, we cannot directly observe the delay on the B-C link. Observation suggests that delays for successive packets are almost independent of one another; in particular the measured delays for two packets sent 20ms apart, the first from A to B (and return), the second from A to B to C (and return), are almost independent. We model this situation by assuming there are distributions $F_X$ and $F_Y$ that give the delays on the links A-B and B-C respectively, with the distribution of the A-C delay being the convolution of these two distributions. In practice we are interested in identifying changes in the distributions as rapidly as possible. However a more basic question is, how to estimate the distribution $F_Y$ when we can observe only $X$ and $Z$?

While our formulation of the deconvolution problem seems natural in our context, we have not seen any study of it in the literature. A Google Scholar search for titles containing "deconvolution" yields about 12000 references; many of these refer to "blind deconvolution" which is what a statistician would term "estimation of a transfer function". If we delete titles containing "blind" there remain about 4730 titles. Most of these are in various applied journals, relating to a large variety of disciplines. A selection of those in statistical and related journals are listed in the References section. In all the papers we have seen, the distribution of $X$ is assumed known.

## 2. A note on notation

The usual convention is to write all mathematical variables in italics, with random variables in upper-case, and realizations in lower-case. We depart from this by using typewriter font `like this` for both observations and functions of them. Our algorithms are copied directly from implementations in the S language. Most of the S functions that we use are self-explanatory, but a detailed explanation appears in Appendix 1. Only two things need explanation here; the function `c()` (concatenate) makes its arguments into a vector. Also, many S functions take a vector argument. It is convenient that subscripts are not used in S; indices are shown by using square brackets. Thus a vector `x` of length 3 has elements `x[1],x[2],x[3]`. This notation makes it easy to write complicated expressions as indices.

[1]Avaya Labs Basking Ridge, NJ, USA, e-mail: `colinm@avaya.com`

### 3. Two naive methods, and a new idea

Recall that the observed samples are `x = c(x[1],...,x[m])` and `z = c(z[1],...,z[n])`. If we have `m = n`, a first suggestion is to sort `x` and `z`, forming `sortx` and `sortz`, and to form `yhat ← sortz − sortx` (i.e. `yhat[i] = sortz[i] - sortx[i]`). If the distributions of $X$ and $Z$ are Normal with variances $\sigma^2$ and $\tau^2$ respectively, so that what we want is an estimate of a Normal distribution with variance $\tau^2 - \sigma^2$, this method produces an estimate of a Normal distribution with the correct mean but with variance $(\tau - \sigma)^2$ (because the sorted vectors are perfectly correlated), which is too small. The method is not consistent as $n \to \infty$.

Another approach, still assuming `m=n`, is to put both `x` and `z` into random orders and to compute the vector of differences `z-x`. Again, this does not work; this gives an estimate of the distribution of $X + Y - X'$ where $X'$ is an independent copy of $X$. In the Normal case described above, this method gives an estimate of a Normal distribution with variance $\tau^2 + \sigma^2$ instead of $\tau^2 - \sigma^2$.

The new idea is that a useful estimate could be obtained if we knew the "right" order in which to take the `zs` before subtracting the `xs`; and we can estimate an appropriate order by a simulation. Here is a first version of how this would work, assuming `m=n`. Suppose we have a first estimate of $F_Y$, represented by a vector of values `oldy = c(oldy[1],...,oldy[n])`. We choose a random permutation `rperm` of $(1,\ldots,n)$, and put the elements of `oldy` into this order. We add the `xs` to give a vector `w` where

$$\mathtt{w} \leftarrow \mathtt{x} + \mathtt{oldy[rperm]}$$

We record the ranks of the elements of this vector. We put the elements of `z` into this same order and subtract the `xs`. Thus

$$\mathtt{newy} \leftarrow \mathtt{sort(z)[rank(w)]} - \mathtt{x}$$

We can repeat this operation as many times as we like.

We will attempt an explanation of why this might be expected to work below. An example is shown in Figure 1. Here the sample size is `n = 100`, and both $X$ and $Y$ are standard Normal. We generated pseudo-random samples `z0 = x0 + y0` and `x1`, placed these in sorted order (`sortz0 = sort(z0)` and `sortx1 = sort(x1)`) and started the algorithm by taking `y1 = sort(sortz0 - sortx1)`. Successive versions of `y` were obtained using the iteration. Note that `rank(runif(n))` is a random permutation of $(1,\ldots,n)$.

$$\mathtt{newy} \leftarrow \mathtt{sort(sortz0[rank(sortx1 + oldy[rank(runif(n))])])} - \mathtt{sortx1}.$$

We ran the iteration for 100 steps. Figure 1 shows the first nine `y` vectors, each sorted into increasing order, plotted against standard normal quantiles. Also shown is the straight line that corresponds to a normal distribution with mean `mean(z0)`

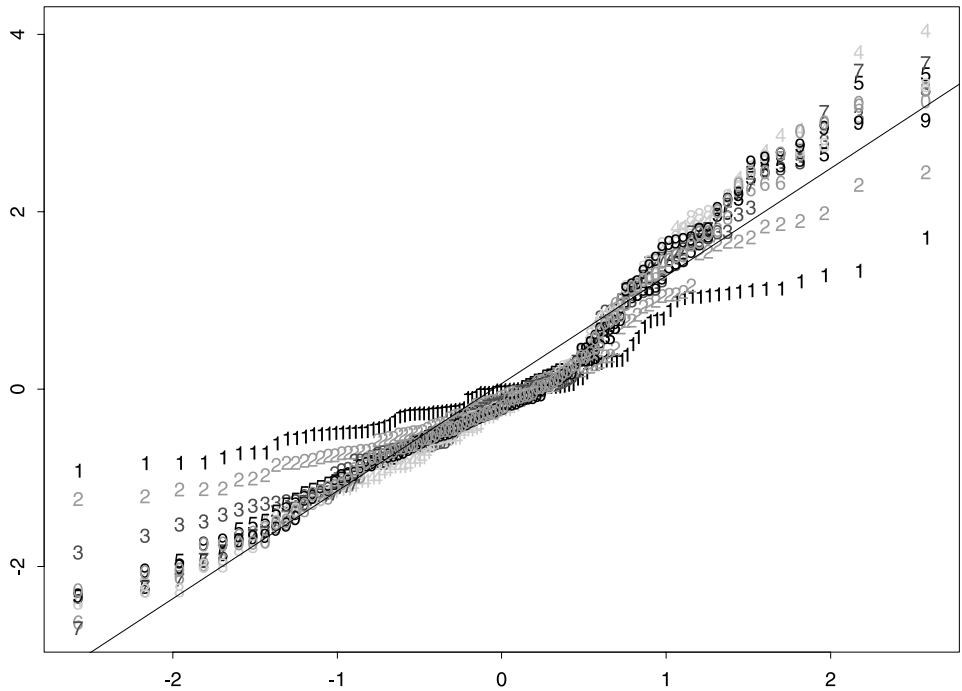

Fig 1. *QQplots of the first nine iterations for the Normal example.*

- `mean(x1)` and variance `var(z0) - var(x1)`. Figure 2 shows iterations 81:100. Figure 3 shows values of a distance index `d`, which is the sum of absolute vertical deviations between this line and the estimate `y`. The algorithm appears to be stable, meaning that in repeated applications of the algorithm, the estimates stay close together. The initial transient takes no more than four iterations. The average value of the distance `d` over iterations 5:100 is 19.69. Also shown (with plotting character "o") are comparable values of `d` for random samples from a normal distribution with the same mean and variance as this fitted normal distribution. The average value of these is 14.74. If we average the `y` vectors over iterations 5:100, we get a vector whose distance from this fitted normal distribution is `d` = 17.82. The average distance between the iterates and their average is `d` = 8.95. Thus the average distance between the iterates and their average is smaller than the average distance between random normal samples and the population line.

The iteration seems to be giving good estimates of $Y$. Why should this be so? Here is an argument to support this expectation. Suppose `z = x+y`; these vectors are realizations of random variables $X, Y, Z$. we cannot observe any of `x,y,z` but can see `sortz = sort(z)` and an independent realization of $X$, namely `x1`. How can we define an estimate of `y`? Since $X$ and $Y$ are independent (by assumption), if `rperm` is a random permutation, then `zhat = sort(x) + sort(y)[rperm]` is a realization of $Z$, sorted according to `sort(x)`. To retrieve `y` we simply subtract `sort(x)` from `zhat`. If `n` is large, we expect `zhat` to be close to `z`, and `sort(x1)` to be close to `sort(x0)`. Thus we expect that putting `z` into the same order as `zhat` will make `z` approximately equal to `zhat`; and subtracting `sort(x1)` from this will approximately retrieve `y`. This argument does not explain why the iteration should converge when it is started with `y0` remote from the correct value. We do not yet

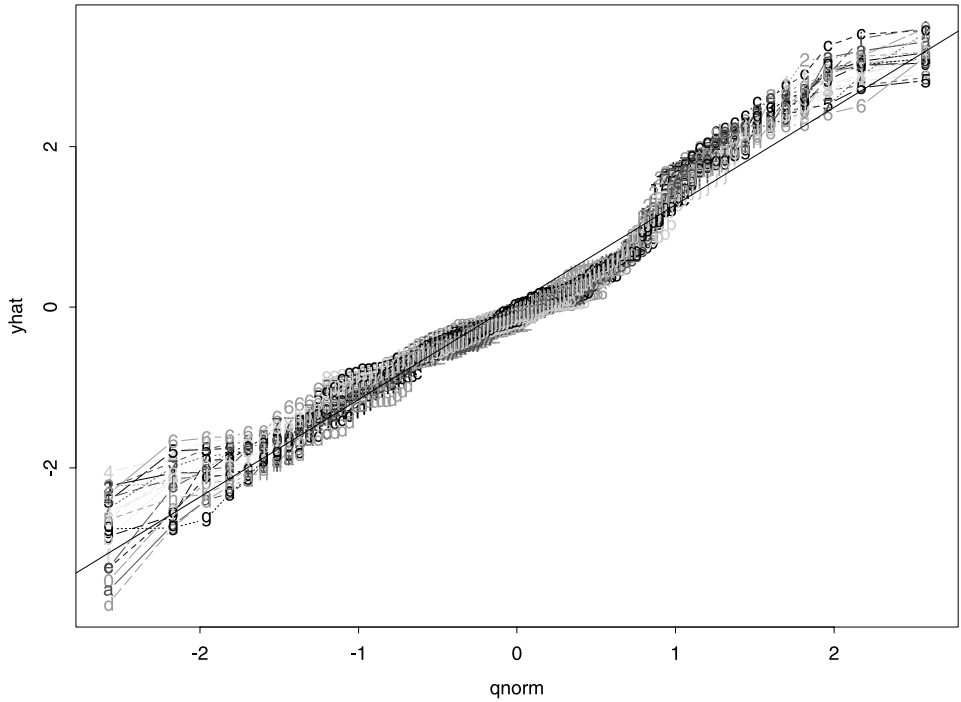

Fig 2. *QQplots of iterations 81:100 for the Normal example.*

have an explanation of this.

## 4. Questions

Several questions come to mind immediately. Is this algorithm always stable? Is the algorithm consistent, meaning that as $\mathtt{n} \leftarrow \infty$, the empirical c.d.f of $\mathtt{y}$ converges in probability to $F_Y$? I thank a referee for reminding me that $F_Y$ may not be unique. What happens when it is not?

To approach these questions, we point out that in the algorithm we have described, the possible values of the vector $\mathtt{y}$ are all of the form $\mathtt{z[perm]} - \mathtt{x}$ where $\mathtt{perm}$ is a permutation of $(1,\ldots,\mathtt{n})$. Thus in repeated applications $\mathtt{y}$ executes a random walk on the $\mathtt{n}!$ possible values of this vector. This random walk will have a stationary distribution, which may not concentrate on a single state (this seems to be the usual case). Some states may be transient. Thus the most we can hope for is that this stationary distribution is close to $F_Y$ in some sense.

Clearly we need a proof that as $\mathtt{n} \to \infty$ this stationary distribution converges (in some sense) to a distribution that is $F_Y$ whenever this is identifiable. Also it would be very pleasant to understand the distribution of the discrepancy measure $\mathtt{d}$ when $\mathtt{y}$ is drawn from the stationary distribution. As yet we do not have these results, but empirical evidence strongly suggests that the convergence result holds universally, and that useful estimates are obtained in all cases. However the dispersion among successive realizations of $\mathtt{y}$ is an over-optimistic estimate of the precision of the estimate of $F_Y$.

Detailed analysis of the stationary distribution seems out of reach. Even with $\mathtt{m} = \mathtt{n} = 3$, 924 different configurations of $\mathtt{x}$ and $\mathtt{z}$ need to be considered. There are

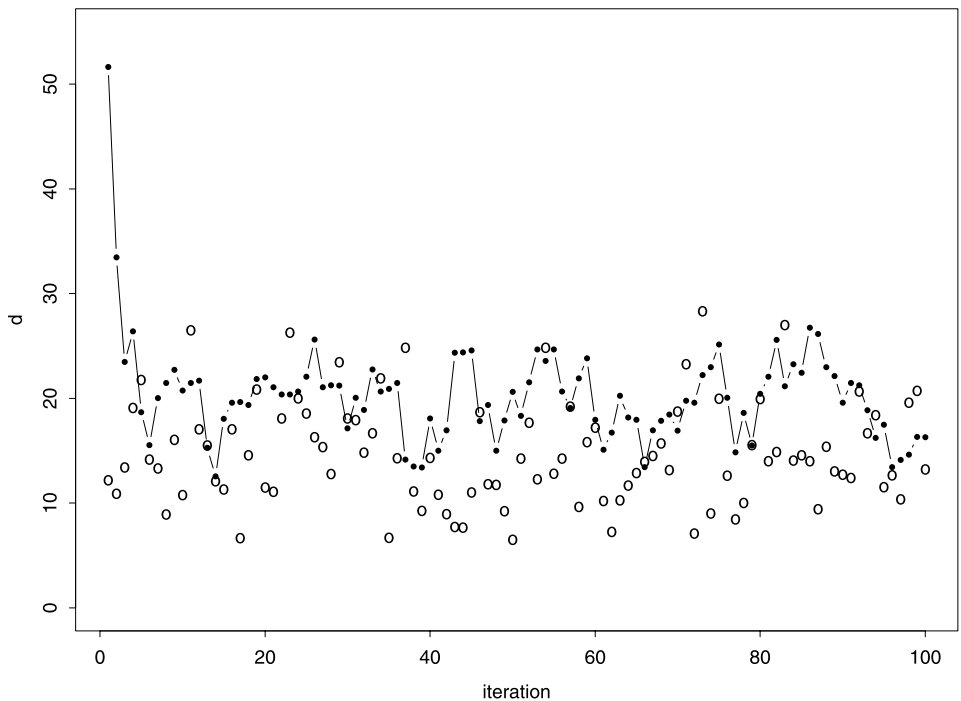

Fig 3. *The index d for the first 100 iterations, with values for random normal samples.*

208 distinct stationary distributions. See Appendix 2.

We suggest that in practice we need to ignore an initial transient, and that the dispersion among successive realizations of y is an over-optimistic estimate of the precision of the estimate of $F_Y$.

We need to consider how to handle boundary conditions, for example (as in the motivating example) that all values of $Y$ are positive. The algorithm as stated need not generate vectors y that satisfy such conditions. Also, we question how the algorithm will perform when there are remote outliers in either or both z0 and x1. Since these samples are assumed to be independent of one another, there is no reason to hope that subtracting an x1 outlier from a z0 outlier will make any sense. We study these questions in Section 6 below.

## 5. Variations

Several variations on the basic idea are as follows.

(a) Instead of using the actual data (x[1],...,x[n]) use a sample from an estimate of $F_X$, for example a bootstrap sample from the observed x.
(b) To add some smoothness to the algorithm, at each iteration replace x by $x + \xi$ and/or y by $y + \eta$, where $\xi$ and $\eta$ are vectors of small Gaussian perturbations. We can use the same perturbations throughout, or we can use independent perturbations at each step of the algorithm.
(c) Similarly we can (independently) smooth z by adding $\zeta$. If we arrange that $\mathrm{var}\zeta = \mathrm{var}\xi + \mathrm{var}\eta$, these smoothings should not introduce any bias into the estimate of $F_Y$, because $X + \xi + Y + \eta$ is distributed like $Z + \zeta$. Of course the efficiency of the method will degrade if the variance of $\zeta$ becomes large (unless each of $X, Y, Z$ is Gaussian).

If `m` and `n` are not equal, to apply the algorithm we need to generate equal numbers of values of `x` and `z`. We can do this either by

 (d) creating vectors of some length `N` by bootstrapping from the observed `x` and `z` (`N` could be very large, so that we are effectively regarding `x` and `z` as defining empirical distributions),
 (e) if `m > n`, by taking `z` with a random sample (without replacement) from `x`; or similarly sampling `z` if `n > m`; or
 (f) if `m<n`, suppose `n = km+r` with `r<m`. Then generate `n` values of `x` by repeating `x` `k` times and adjoining a random sample of size `r` drawn from `x`. Similarly if `m > n`, repeat `z` to fill out `m` values.
 (g) In generating `w` we can use a bootstrap sample from `y`, possibly smoothed as above.

To achieve stability in the estimate of $F_Y$, we can

 (h) Apply the algorithm a moderate number of times, `k` say, and average the resulting sorted `y` vectors; or
 (i) concatenate successive `y` vectors to form a pooled estimate of $F_Y$; if we do this we can at each stage
 (j) generate `w` by sampling from this pooled estimate.

It is not clear how to generalize the idea to deal with multivariate observations.

## 6. Boundary conditions, and outliers

If some bound on $Y$ is known a priori, for example if it is known (as in the motivating problem) that $Y > 0$, we need to decide what to do if the algorithm produces one or more negative values in `y`. Some possibilities in this case are:

 (k) At each iteration, round negative values of `yhat` up to zero.
 (l) At each iteration, replace negative values by randomly sampling from the positive ones;
 (m) At each iteration, replace negative values in `yhat` by copies of the smallest values among the positive ones.
 (n) At each iteration, reject a random permutation if it leads to offending values; draw further permutations until one is obtained that satisfies the positivity conditions;
 (o) At each iteration, adjust the permutation by changing (at random) a few elements (as few as possible?) in such a way as to meet the conditions.

Our experience so far suggests that none of these proposals works very well. Proposals (n) and (o) are excessively tedious, and have been tried only in very small examples. At this point we recommend another strategy, namely

 (p) Replace negative values in `yhat` by their absolute values.

We investigated two of these proposals as follows. We generated 100 pseudo-random exponential variates `x0`, and added a similar (independent) vector `y0` to form the observed vector `z0`. We assumed that an independent vector `x1` was also observed. We ran the iteration in three ways:

 (q) no adjustment

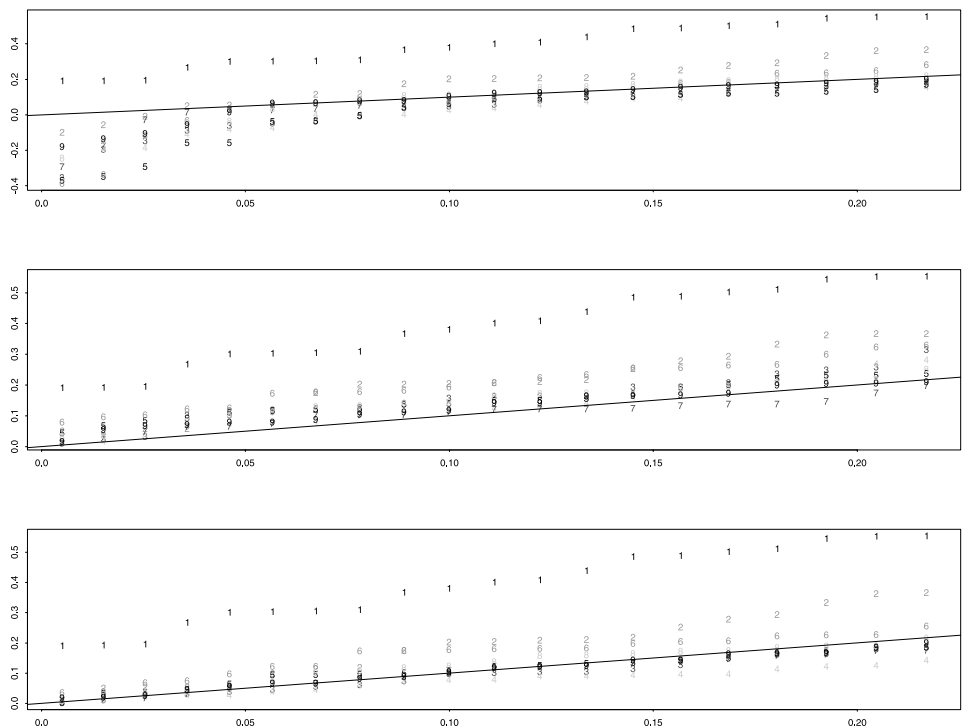

FIG 4. *The lowest 20 elements of the first nine iterations for each of three methods: Top:(q), Middle:(l), Bottom:(p).*

(l) replace negative values by a random sample from the positive values;

(p) Remove negative values of `yhat`, replacing them by their absolute values. This can be done in S by an application of the `abs` function:

$$\mathtt{newy} \leftarrow \mathtt{sort(abs(sortz0[rank(sortx1} \\ +\mathtt{oldy[rank(runif(n))])] - sortx1))}.$$

All three methods performed similarly for values of `yhat` greater than 0.25.

Figure 4 shows the lowest 20 values of `yhat` for the first nine iterations, plotted against standard exponential quantiles, for each of these three methods, together with the line through the origin with slope `mean(z0) - mean(x1)`. We see that the naive method (q) produces a large number of negative values; method (l) avoids this but seems to introduce a positive bias; method (p) works well.

Figure 5 shows the number of negative elements in `yhat` (before adjustment) for the three methods. The average numbers over the first 100 iterations are (q) 4.44, (l) 3.35, (p) 2.68. We have no understanding why the "absolute values" method works as well as it seems to.

We have run similar trials for the case where both $X$ and $Y$ are uniform on $(0, 1)$, so that $Z$ has a triangular density supported on $(0, 2)$. Here for method (p) we need to reflect values above `y=1` to lie in $(0, 1)$. Again, method (p) seems to be better than (q) and (l).

We have also investigated the performance of the "absolute values" method when there is a positivity condition and outliers are present. We find that the non-outlying part of the distribution is estimated satisfactorily. We took `x0`, `x1`, and `y0` each to contain 95 samples from a standard exponential distribution (with mean 1), and 5 samples from an exponential distribution with mean 100. Figure 6 shows the first four iterations of our basic algorithm, using the option (p) to adjust negative estimates, plotted against `sort(y0)`. Figure 7 expands the lower corner of this plot, with the line through the origin of unit slope. The iterates seem to be staying close to this line.

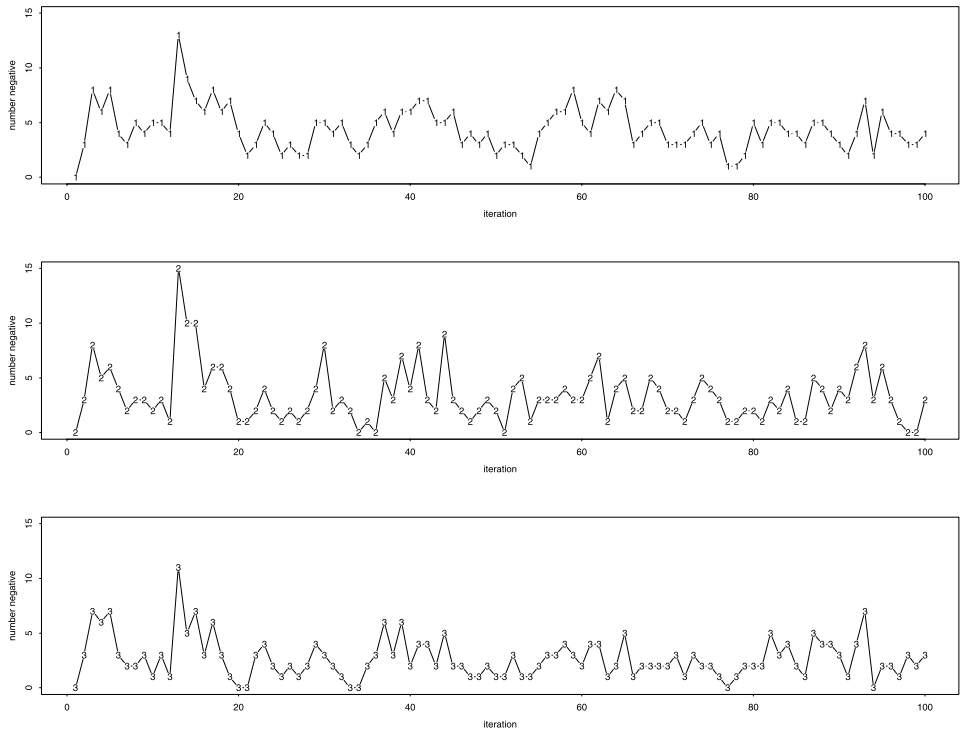

Fig 5. *The number of negative elements (before adjustment) for each of three methods: Top:(q), Middle:(l), Bottom:(p).*

At this point our recommendation (if $m = n$ and the variables are continuous, so that there are no ties in the computed values), is to use the original method, i.e. do not bootstrap or smooth or use (j). If the variables are lattice-valued, for example integer-valued, it seems to help to add small random perturbations to `x` and `y[rperm]` at each stage to break the ties randomly. It is not clear whether it is as good to simply add small perturbations once and for all. To handle the boundary and outlier problems, we recommend using the absolute-values method (p) above.

**Appendix 1. The S language**

In S the basic units of discourse are vectors; most functions take vector arguments. The elements of a vector `x` of length `n` are $x[1], \ldots, x[n]$. `c()` is the "concatenate" function, which creates a vector from its arguments. Thus $x = c(x[1], \ldots, x[n])$. If the elements of an `m`-vector `s` are drawn from $1, \ldots, n$, (possibly with repetitions), `x[s]` is the vector $c(x[s[1]], \ldots, x[s[m]])$. The function `sort()` rearranges the elements of its argument into increasing order; so if $x = c(2,6,3,4)$, `sort(x)` is $c(2,3,4,6)$. The function `rank()` returns the ranks of the elements of its argu-ment; i.e. `rank(x)[i]` is the rank of `x[i]` in x. Thus if `x = c(2,6,3,4)`, `rank(x)` is `c(1,4,2,3)`. `sort(x)[rank(x)]` is just x. Another function we have used is `runif()`, which generates pseudo-random uniform variables drawn from the interval (0,1). Thus `rank(runif(n))` is a random permutation of $1, \ldots, n$ `rnorm(n)` generates n standard normals; `rexp(n)` generates n random exponentials. The function `abs()` replaces the elements of its argument by their absolute values.

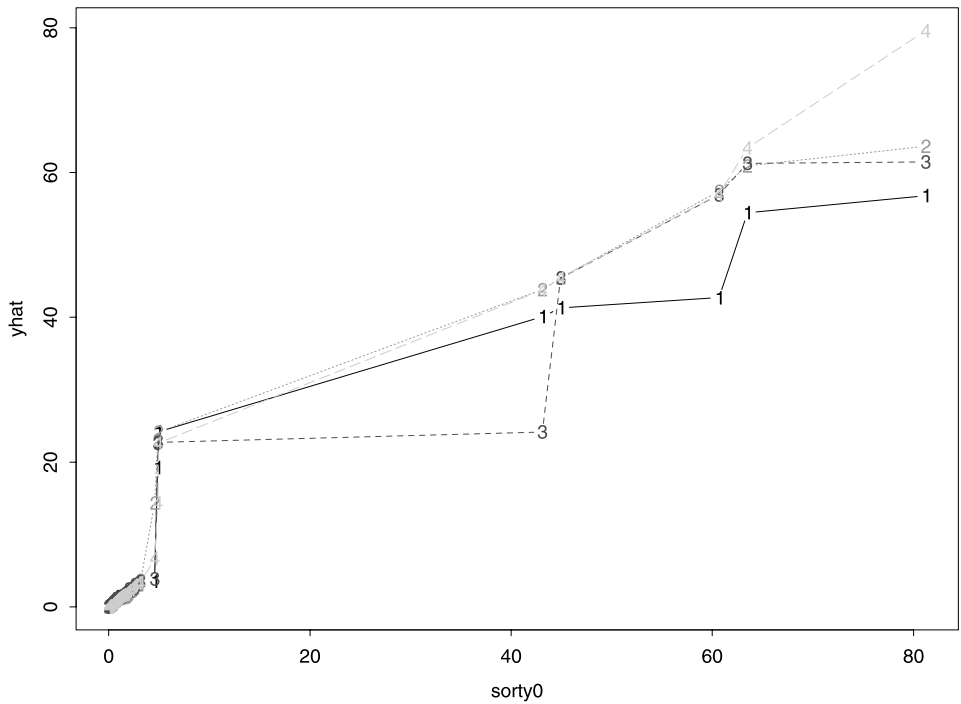

Fig 6. *Four iterates of the basic algorithm, using option (p) to handle the positivity condition, when outliers are present.*

**Appendix 2. The case `m=n=3`**

Without loss of generality we may assume `x1 = c(0,x,1)` and `z0 = c(-a,0,b)` with `0 < x < 1/2` and `a` and `b` positive. Examination of the 36 possible values of `(z0[perm1]-x1)[perm2] +x1` shows that the stationary distribution will change whenever any of `a,b` and `a+b` crosses any of the values `x,2x,1,1+x,1-x,1-2x,2, 2-x,2-2x`. For a general x in (0,1/2) these cut-lines divide the positive quadrant of the `a,b` plane into 154 regions. The configuration of these regions changes when x passes through the values (1/6,1/5,1/4,1/3,2/5). Thus we need to consider six representative values of x, perhaps `x = c(10,22,27,35,44,54)/120`, and for each of these values of x we have 154 regions, 924 regions in all. We computed the transition matrix of the random walk for each of these 924 cases, and found 208 different stationary distributions. One of these, where one state is absorbing and the other five transient, occurs 84 times. Ten distributions occur only once each.

10    C. Mallows

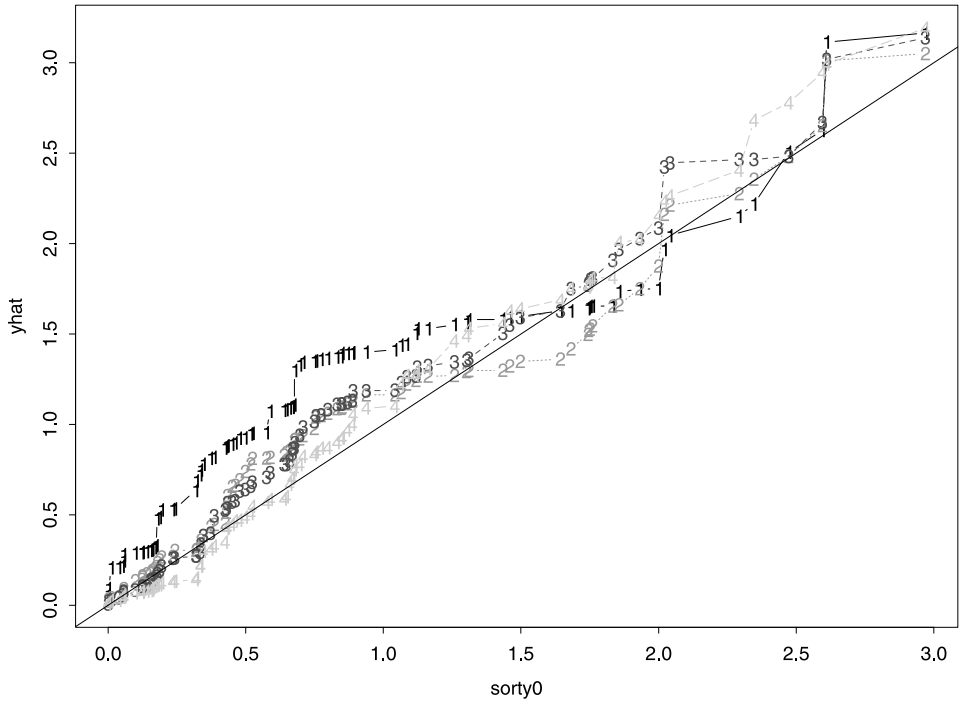

Fig 7. *Expansion of the lower corner of Figure 6.*

A similar calculation for `m` or `n` larger than 3 seems impractical.
**Acknowledgments.** Thanks to Lorraine Denby, for showing me the problem, and to Lingsong Zhang, who did some of the early simulations. Also to Jim Landwehr, Jon Bentley and Aiyou Chen for stimulating comments. Two referees contributed insightful remarks.